\documentclass[prd,preprint,eqsecnum,nofootinbib,amsmath,amssymb,
               tightenlines,dvips]{revtex4}
\usepackage{graphicx}
\usepackage{epsfig}
\usepackage{bm}

\begin {document}


\def\Mrowczynski{Mr\'owczy\'nski}

\def\Eq#1{Eq.\ (\ref{#1})}
\def\alphas{\alpha_{\rm s}}
\def\NCS{N_{\rm CS}}

\def\D{{\bm D}}

\def\k{{\bm k}}
\def\half{{\textstyle{\frac12}}}

\def\p{{\bm p}}

\def\x{{\bm x}}

\def\v{{\bm v}}
\def\E{{\bm E}}
\def\B{{\bm B}}
\def\A{{\bm A}}

\def\grad{{\bm\nabla}}

\def\tr{\operatorname{tr}}

\def\lmax{\ell_{\rm max}}
\def\ldamp{\ell_{\rm damp}}



\title
    {
    QCD Plasma Instabilities:  The Nonabelian Cascade
    }

\author{Peter Arnold}
\affiliation
    {%
    Department of Physics,
    University of Virginia, Box 400714,
    Charlottesville, Virginia 22901, USA
    }%
\author{Guy D. Moore}
\affiliation
    {%
    Department of Physics,
    McGill University, 3600 rue University,
    Montr\'eal QC H3A 2T8, Canada
    }%

\date {September 20, 2005}

\begin {abstract}%
    {%
       Magnetic plasma instabilities appear to play an important role in the
       early stages of quark-gluon plasma equilibration in the
       high energy (weak coupling) limit.
       Numerical studies of the growth of such instabilities from
       small seed fluctuations have found initial exponential
       growth in their energy, followed by linear growth once the
       associated color magnetic fields become so large that their
       non-abelian interaction are non-perturbative.
       In this paper, we use simulations to determine the nature of
       this linear energy growth.  We find that the long-wavelength
       modes associated with the instability have ceased to grow, but
       that they cascade energy towards the ultraviolet in the form of
       plasmon excitations of ever increasing energy.
       We find a
       quasi-steady-state power-law distribution $f_k \propto k^{-\nu}$
       for this cascade, with spectral index $\nu \simeq 2$.
    }%
\end {abstract}

\maketitle
\thispagestyle {empty}


\section {Introduction}
\label{sec:intro}

One of the surprises which has emerged from the heavy ion experiments at
RHIC is that the medium or plasma which is produced in a heavy ion
collision appears to display collective behavior analogous to a fluid
with a very small viscosity (the so-called ``Quark-Gluon Liquid'')
\cite{several_papers}.  In particular, hydrodynamic treatments taking
the medium to be an ideal fluid give the correct flow description
\cite{hydro_results}.  To clarify, we remind the reader that ``ideal
fluid'' is the opposite of ``ideal gas''; it means that scatterings or
interactions so efficiently maintain local equilibrium (or at least
isotropy), that the stress tensor remains everywhere isotropic when
measured in the local rest frame.  Gases behave like ideal fluids on distance
and time scales large compared to transport mean free paths and mean
free times.

The most popular, and perhaps most likely, explanation for the small
viscosity is
that the Quark-Gluon Plasma at the energy densities achieved at RHIC is
very far from weak coupling.  The strong coupling $\alphas$ is
large, interactions are very strong, and the collective behavior is
natural.  To secure this interpretation, we would need to see that the
observed behavior
really differs from the expected weakly coupled behavior.  Most
treatments to date of weakly coupled hot QCD show that equilibration
is slow \cite{slow_eq} (see however \cite{greiner}).  However, we have
recently shown \cite{ALM} that even the most complete of these, the
``bottom-up'' scenario of Baier, Mueller, Schiff and Son
\cite{bottom_up}, is incorrect, because it ignores plasma instabilities
\cite{plasma_old}, which in fact dominate the physics of
weakly-coupled anisotropic plasmas, in QCD or in ordinary QED.

\begin{figure}[b]
\centerline{\epsfxsize=0.9\textwidth \epsfbox{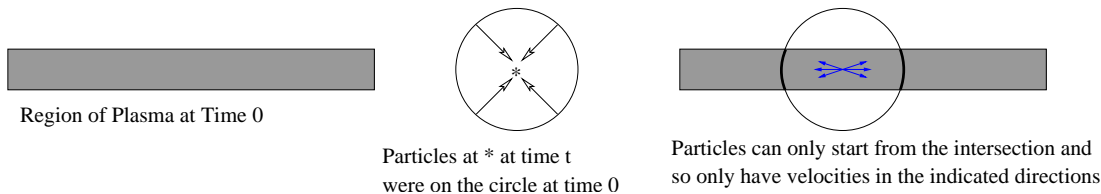}}
\caption{\label{fig1} Cartoon of why the momentum distribution becomes
anisotropic.  The starting region is thin along the beam axis (vertical
in this figure).  Since particles fly in a straight line at the speed of
light, particles at point $*$ at time $t$ originated at time 0 on a
sphere of radius $t$ centered at $*$.  But only part of this sphere was
in the plasma; so only particles with small beam-axis momenta can get to
point $*$ at time $t$, and the momentum distribution is anisotropic.}
\end{figure}

Plasma instabilities arise as a result of two pieces of physics.  
First, Lorentz contraction of the nuclei means that the initial region
of plasma is a flat pancake shape.  It is reasonable to expect at weak
coupling that quarks and gluons will for a time fly in nearly straight lines at
near the speed of light (at least, once the system has expanded enough
that the densities of quarks and gluons are perturbative).
Consider particles scattered in all directions in the initial moments of
the collision.
The starting geometry dictates that the
momentum distribution of particles will subsequently
become highly anisotropic with time, as
illustrated in Fig.\ \ref{fig1}.
Second, in the presence of such anisotropic momentum distributions,
certain soft gauge field configurations grow exponentially, at least if
they are small enough for a perturbative treatment to be reliable.

Not only have plasma instabilities not been fully taken into account in
studying the evolution of the Quark-Gluon Plasma;
instabilities in
Yang-Mills theory in general are not yet well understood or well
characterized. 
Therefore, at the current juncture we cannot say whether the
experimental results at RHIC are in contradiction with weakly coupled
predictions, because we simply do not know what the predictions of
isotropization in a weakly coupled Quark-Gluon Plasma actually are.
This rather embarrassing gap in our knowledge needs to be filled.
In particular, the behavior of
plasma instabilities in a nonabelian theory such as QCD are expected to
be intrinsically different from the case of an ordinary QED plasma,
because nonabelian gauge fields can directly interact with each other,
and because charge carriers can have their colors rotated by intervening
QCD fields.

Recently there has been substantial progress on this problem, though it
is far from settled.  The growth of QCD plasma instabilities when the
soft colored gauge fields are small has been well characterized
\cite{RS,ALM}. Arnold and Lenaghan \cite{AL} conjectured that
exponential growth would continue after QCD fields became large.
1+1 dimensional simulations \cite{RRS,Nara} confirmed this behavior, but
3+1 dimensional studies \cite{linear1,RRS2} showed something different: 
when soft colored gauge fields become large due to instability growth,
exponential growth shuts off and the energy in soft
gauge fields instead grows {\it linearly}\/ with time.
Recent partial attempts to explore the consequences of instabilities
for thermalization may be found in Refs.\ \cite{newBUP}.

The goal of this paper is to further investigate the behavior of an
anisotropic quark-gluon plasma, after soft gauge fields have become
large.  We will show that the soft gauge fields with wave-numbers which
should make them unstable instead enter a dynamical
quasi-steady state, gaining energy from the instability but losing
energy, via their nonabelian interactions, to more ultraviolet field
excitations.  The energy released into infrared gauge fields by the
instability then cascades towards higher wave-number gauge field modes,
with occupation number $f(k) \propto k^{-\nu}$
up to a time-dependent cutoff $k_{\rm max}(t)$.
Through simulations, we measure the spectral index to be $\nu \simeq 2$.


\section{Setup and formalism}
\label{sec:setup}

We will carry out a numerical study of nonabelian, classical Yang-Mills
theory coupled to an anisotropic bath of high momentum particles.  The
numerical tools were already presented in detail in \cite{linear1}, and
this paper is a continuation of that work.  To make our presentation more
self-contained, we will briefly review the setup and formalism here.

Our treatment is founded on two assumptions about the system under
study.  The first is that the coupling $\alphas \ll 1$.  The second is
that there is a separation of scales between the momentum of the typical
excitation, $p$, and the {\sl screening scale} $k \ll p$.  The screening
scale is set by the momentum $p$ and number density $n$ of the typical
excitations as $k^2 \sim g^2 n/p$.  The number density is given by
$n = \int_\p f_\p \sim p^3 \bar f_p$, where $f_\p$ is the occupancy of
momentum state $\p$ and $\bar f_p$ is its angular average.
Therefore there is a scale separation $k \ll p$
provided that the angular-averaged occupancy $\bar f_p$
of typical excitations is $\ll 1/g^2$.
This is essentially a diluteness condition
on the typical excitations in the system.

In the context of heavy ion collisions, the condition $\alphas \ll 1$ is
formally valid for the collision of extremely large and high-energy
heavy ions \cite{MV_maybe,bottom_up}.
The initial angular-averaged occupancy is at most $\bar f_p \sim 1/g^2$,
which is the saturation limit \cite{MV_maybe,bottom_up}%
\footnote{
    If the initial occupancies were $\gg 1/g^2$, then the magnetic
    screening length they would provide would be larger than the imputed
    typical momentum.  However, the concept of momentum for an
    excitation of a gauge dependent field is only well defined and
    robust under gauge fixing prescriptions to an accuracy set by the
    magnetic screening length.
    These ideas are central to the idea of
    the saturation scale.  However this argument does not depend on the
    saturation scenario as such.
}
(corresponding to number and energy densities $n \sim p^3/g^2$
and $\epsilon \sim p^4/g^2$).
The density $n$ of the original particles, and so $\bar f_p \sim n/p^3$, will
fall with time as the system expands.
So the scale separation condition $\bar f_p \ll 1/g^2$
will be satisfied at least for
$\tau \gg \tau_0$
(up until the much later time when the particles of momentum
$p$ lose their energy and the plasma thermalizes---see, for example,
the discussion in the original bottom-up scenario
of Ref. \cite{bottom_up}).

If we have a clean separation between the screening scale $k$ and
typical momenta $p$, then we can describe the physics on length scales
$x\sim 1/k$ in terms of (i) classical fields with wave vectors $\sim k$
and (ii) a distribution of classical particles (possessing well defined
momenta and positions) representing excitations with momenta $\sim p$.
This is a Vlasov equation treatment.  With a little more work
\cite{iancu,BMR,linear1}, one may work in terms of a classical field
variable $A^a_\mu(\x)$, a variable $W^a(\v,\x)$, which represents the
net (adjoint) color of all particles moving in direction $\v$ at point
$\x$, and a background colorless particle density with angular
distribution $\Omega(\v,\x)$ and polarizability characterized by a
screening mass $m^2$,
\begin{equation}
m^2 = \sum g^2 t_{\rm r} \int \frac{d^3 p}{(2\pi)^3 p^0}f(\p) \, ,
\end{equation}
with the sum over spin and particle type (including anti-particles).%
\footnote{
  In previous work we have referred to $m$ as $m_\infty$, but here we
  will drop the subscript.
}
Here $t_{\rm r}$ is a color group factor [defined in terms of color
generators $T$ by
$\tr(T^{\rm a} T^{\rm b}) = t_{\rm r} \delta^{ab}$ and
equal to 3 for gluons].
For an isotropic medium, $m^2=\frac{1}{2} m_{\rm D}^2 = \frac{3}{2}
\omega_{\rm pl}^2$,
where $m_{\rm D}$ is the Debye mass and $\omega_{\rm pl}$ is the
plasma frequency.%

We argued in \cite{ALM} that the intrinsic length and time scales for
the instability to develop are short compared to the length and time
scales on which the heavy ion system as a whole evolves.  Therefore we
will restrict our attention in this (still somewhat exploratory) study
to a non-expanding system with spatially uniform $\Omega(\v)$.
The
evolution equations for $A_\mu$ and $W$ are
\begin{eqnarray}
\label{eq:YangMills}
D_\nu F^{\mu \nu}(\x) & = & \int_\v v^\mu W(\v,\x) \, , \\
(D_t + \v \cdot \D_\x) W & = & m^2 \left[ \E \cdot (2\v-\grad_\v)
	+ \B \cdot ( \v \times \grad_\v ) \right] \Omega(\v) \, .
\label{eq:W}
\end{eqnarray}
The first equation is the Yang-Mills field equation, with $W(\v)$
giving rise to a current.  The second equation, derived in
Ref.\ \cite{linear1}, shows how electric and
magnetic fields can polarize the colorless distribution of particles to
create a net color moving in each direction.
The dynamics of the soft fields is equivalent to
that of hard-loop effective theory \cite{MRS}.

For a fully nonperturbative study of this system, we implement these
equations on a lattice.  The treatment of the classical, Minkowski space
gauge fields on the lattice is the standard one \cite{Ambjorn}.  The
representation of the $W$ fields is developed in \cite{BMR,linear1} and
consists of replacing first derivatives with forward-backwards
(covariantized) differences in \Eq{eq:W}, and expanding the space of
velocities $\v$ in spherical harmonics, truncated at a finite but large
$\lmax$.
In this work we make one additional modification which improves approach
to the $\lmax\to\infty$ limit and so reduces the memory and time
requirements for simulations:
We apply a weak damping term to the large $\ell$ spherical
harmonic components of the $W$ field.
The reason this is an improvement is that very
large $\ell$ excitations of the $W$ field tend to cascade to still
larger $\ell$, due to the $\v \cdot \D$ term in \Eq{eq:W}.%
\footnote{
  The $\v\cdot\D$ term mixes neighboring $\ell$'s because
  $\langle \ell m | \v | \ell' m' \rangle$ has non-zero cases
  when $\ell' = \ell \pm 1$.
}
With a finite $\lmax$ cutoff, some of this excitation energy ``bounces
off'' the $\lmax$ cutoff and re-enters the infrared.  This effect
becomes smaller as $\lmax$ is raised, but can be largely eliminated with
the damping term we add.  This point is discussed and justified in greater
detail in appendix \ref{sec:damp}.

The goal of this paper is to understand the behavior of a would-be
unstable system after the unstable modes have already grown to have
nonperturbative occupancy.  In the context of a heavy ion collision,
this is probably the only relevant question; since the exponential
growth rate is larger than the inverse system age for all times after
the formation time of the plasma, there has always been time for
unstable gauge field modes to grow to nonperturbative size.  Therefore
we will begin our system with initial conditions for which the infrared
(IR) fields
are already nonperturbatively large.  Again, since this is a qualitative
exploratory work, we will work in SU(2) rather than SU(3) gauge theory
and we will only consider a single anisotropic particle distribution,
the same as the one considered in Ref.\ \cite{linear1}.

In addition to our focus on large rather than small initial conditions,
there are slight technical differences in our choice of initial
conditions compared to those of Ref.\ \cite{linear1}.
We start the system with
vanishing $E$ and $W$ fields and with the $A$ field selected from the
thermal ensemble at temperature $T=2m/g^2$, but then field-smeared to a
scale $1/2m$.%
\footnote{
  In three dimensions, there should be no important qualitative difference
  between using a smeared thermal distribution for the initial
  conditions, as here, or a
  smeared Gaussian noise distribution, as in Ref.\ \cite{linear1}.
  We have checked this for a variety of properties, such as the rate of
  linear growth of energy with time.  The reason we use a different procedure
  here than in Ref.\ \cite{linear1} is inessential, having to do with
  the development of our code.
}
This is a gauge-invariant procedure which corresponds
perturbatively to multiplying a thermal spectrum for
$A(k)$ by $\exp(-k^2/4m^2)$.  We will explain
the field smearing procedure below because it also constitutes one of
our best measurables for understanding, in a gauge-invariant fashion,
what corresponds to infrared phenomena and what to ultraviolet.


\section{Results: behavior of the nonabelian cascade}
\label{sec:results}

\begin{figure}
\centerline{\epsfxsize=0.7\textwidth\epsfbox{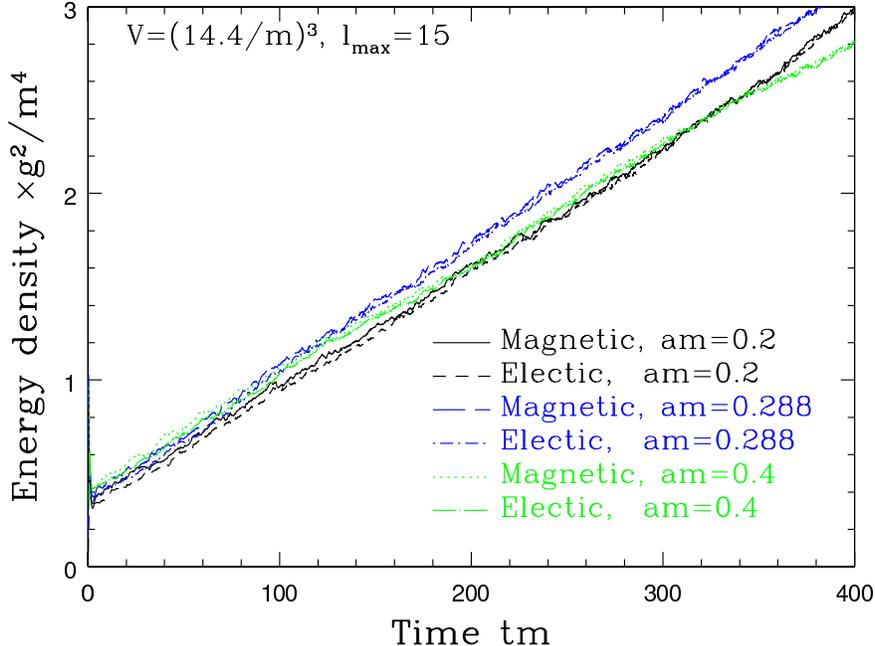}}
\caption{\label{fig:lin} (color online)
Magnetic energy density $\int d^3 x \: B^2/2V$
and electric energy density $\int d^3 x \: E^2/2V$ as a function of
time, for three lattice spacings of $am=0.2$, $0.288$, and $0.4$.  
After a brief transient owing to our choice of initial conditions, the
electromagnetic energy rises linearly with time.}
\end{figure}

Our first conclusion, already presented in our previous paper
\cite{linear1}, is that the energy in soft electromagnetic fields grows
linearly with time.  This is displayed in Fig.~\ref{fig:lin}, which
shows that the linear behavior is common to electric and magnetic fields
and is robust to changes in the lattice spacing.  However, this result
leaves it unclear whether or not this represents continued growth of the
soft unstable fields, whether these fields ``abelianize'' \cite{AL}, and
whether they retain long time scale coherence.  We now present
evidence that the answer to all three questions is, ``no.''

\subsection{Chern-Simons diffusion and IR dynamics}

First, consider the behavior of Chern-Simons number,
\begin{equation}
\NCS(t)-\NCS(0) \equiv \int_0^t dt' \int d^3 x
	\frac{g^2}{8\pi^2} \E^a(x,t') \cdot \B^a(x,t') \, .
\end{equation}
Chern-Simons number is useful because it characterizes nonperturbative
physics.  In an abelian theory, or a nonabelian theory where the fields
are weak, $\NCS$ may fluctuate about zero but cannot
drift away from zero permanently.  That is because permanent change requires
topology change (the Minkowski
version of instantons).  Therefore, ignoring small fluctuations, the time
evolution of $\NCS$ is purely indicative of the dynamics of
nonperturbatively large fields.  Fully topological algorithms for
tracking $\NCS$ in a real-time gauge field evolution already
exist.  We use the one from Ref.\ \cite{broken_sphaleron}, which is a
modification of a technique due to Ambj{\o}rn and Krasnitz
\cite{AmbjornKrasnitz}.

\begin{figure}
\centerline{
\epsfxsize=0.46\textwidth\epsfbox{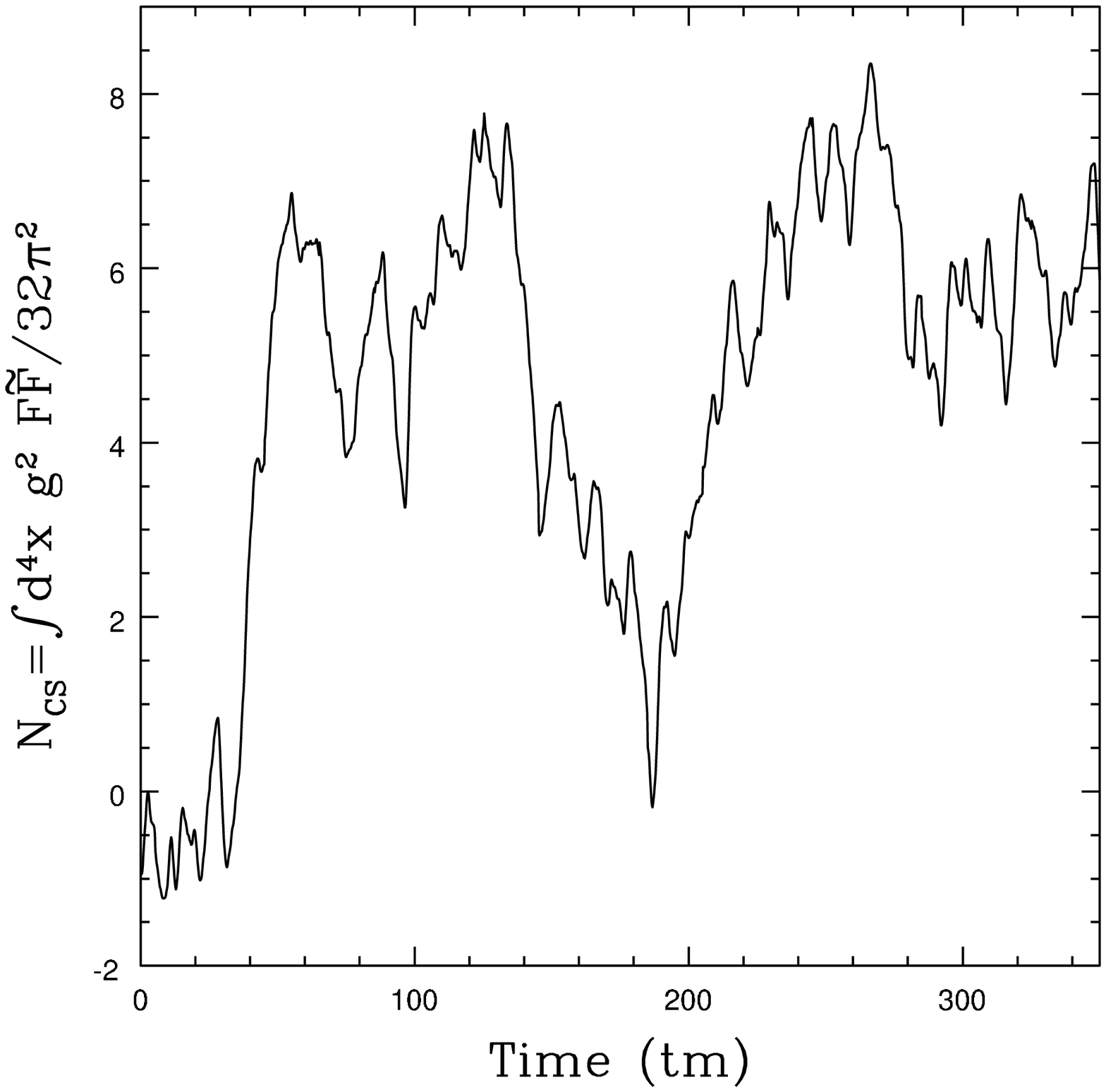}  \hfill
\epsfxsize=0.46\textwidth\epsfbox{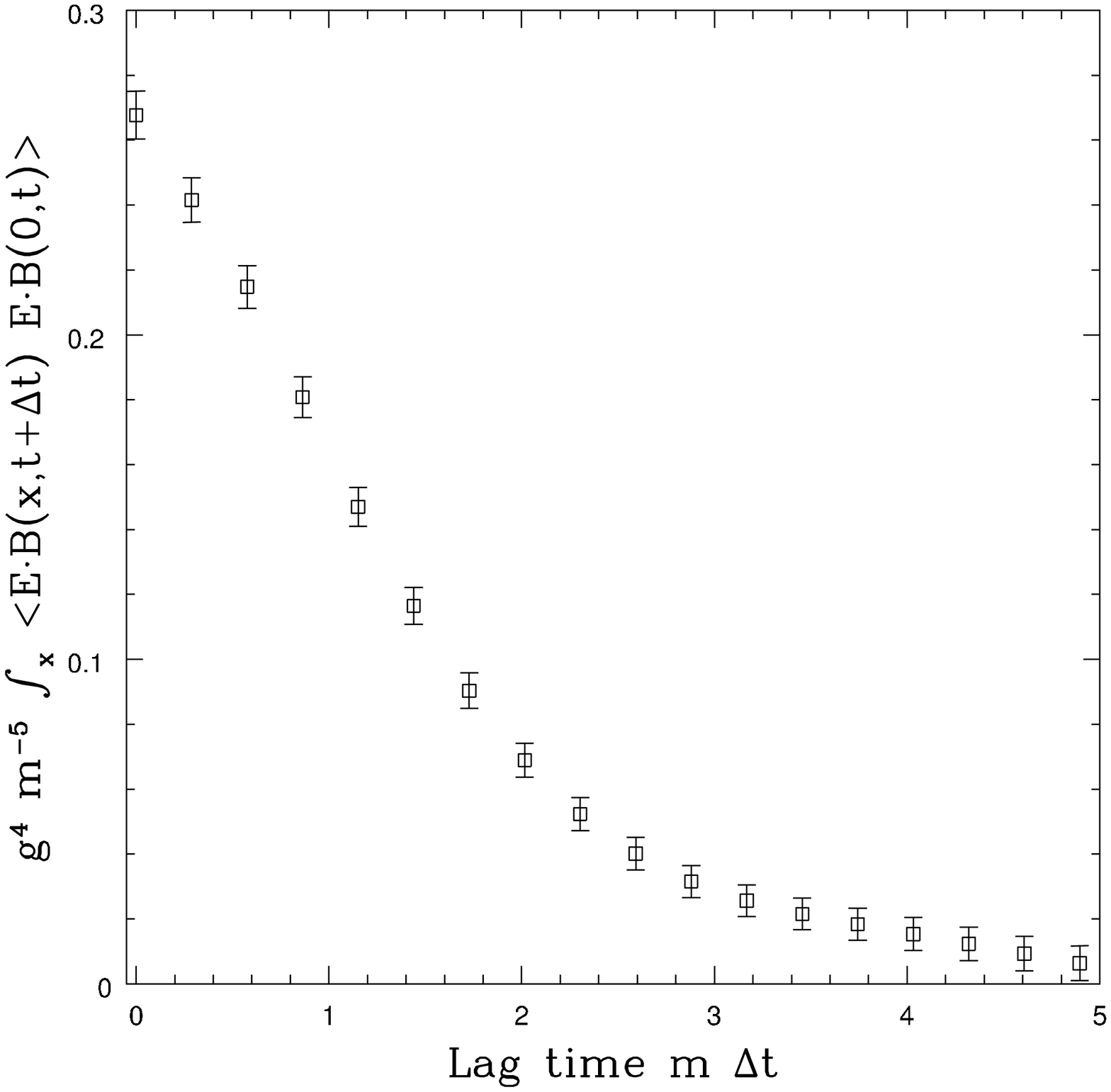}}
\caption{{\em Left:}  typical $\NCS$ evolution in the linear regime.
{\em Right:}  the autocorrelator $\dot{N}_{\rm CS} \dot{N}_{\rm CS}$, or
$\int_x \langle (E\cdot B)[0,t] (E\cdot B)[x,t+\Delta t]\rangle$,
as a function of lag time $\Delta t$.  Chern-Simons
number changes randomly with coherence time scale of order $m^{-1}$.
The measurements were made on spatially
smeared copies of the fields,
as described in \cite{broken_sphaleron}, with smearing extent
$\tau = 0.16/m^2$.
({\em left}:$72^3$ lattice, $am=0.2$, $\lmax=15$;
{\em right}:$50^3$ lattice, $am=0.288$, $\lmax=15$)
\label{fig:NCS}}
\end{figure}

A sample Chern-Simons number trajectory, from the $am=0.2$ evolution
shown in Fig.\ \ref{fig:lin}, is shown on the left in Fig.\
\ref{fig:NCS}.  Chern-Simons number is changing by large amounts
(indicating nonperturbatively large fields), and in a chaotic fashion
(indicating that the fields have no long time-scale stability).  This
means that the soft fields remain large, but evolve dynamically,
changing on a time scale set by the scale $m^{-1}$.  To see this better,
consider the auto-correlator $\langle \dot{N}_{\rm CS}(t+\Delta t) 
\dot{N}_{\rm CS}(t) \rangle$, shown on the right in that figure.  This
correlator indicates over what time scale coherent changes to the gauge
fields occur.  The figure presented represents an average over time and
over 10 independent initial conditions drawn from the same ensemble,
using a $50^3$ lattice with $am=.288$ and $\lmax=15$.  The errors were
determined by the jackknife method.

Although we have not shown it in the figure, we have also checked that
the evolution of $\NCS$ does not speed up or slow down during the course
of the linear rise in magnetic energy.  For instance, comparing the
first and second halves of the evolutions used to make Fig.\
\ref{fig:NCS}, the Chern-Simons number diffusion rate (sphaleron rate)
is consistent between the two halves to within $10\%$ error bars, and
the full auto-correlator is also consistent within errors.
What this tells us is that the IR fields truly are nonperturbative, that
they evolve with a characteristic time scale of order $1/m$, and that
the time scale and the size of the nonperturbative fields is constant
throughout the linearly growing regime.


\subsection{Coulomb gauge spectra}

If IR fields are not growing, then the linear growth of magnetic energy
must reflect growth of higher-momentum modes of the soft gauge field.
To clarify the situation, it would be useful to know
the power spectrum of the soft gauge fields as a function of
wavenumber $k$ and
time $t$.  We will start with a direct but gauge-dependent measurement of
the power spectrum in Coulomb gauge.
Later we will discuss how the spectrum can be
accessed indirectly through gauge-invariant measurement involving
smearing (also known as cooling) of the field configurations.

Our picture (to be supported by data below) is that,
during the linear growth phase of the total energy in soft fields,
the soft fields consist of (i) a non-perturbative IR component plus
(ii) a perturbative component in the form of higher-momentum plasmons.
We would like to know the distribution function $f(k)$
of these plasmons as a function of $k$.
We fix Coulomb gauge
using the standard algorithm \cite{MandulaOgilvie} adopted to the
real-time case \cite{slavefield}.
Then we (i) extract the Fourier spectrum of $\A$ and $\E$, (ii) evaluate the
two point function, averaging over $\k$ vectors in narrow blocks of
$|\k^2|$,\footnote%
    {%
    technically, we use the lattice $\k^2$, $\sum_i (4/a^2)\sin^2
    (k_ia/2)$. 
    }
and (iii) define the distribution function as determined by $\A$ and as
determined by $\E$ through%
\footnote{
  Consider the total energy of a gas of weakly-interacting,
  high-momentum
  ($k \gg m$) plasmons.  This could be written in terms of $f$ as
  $N_{\rm dof} V \int_\k \omega_k f_\k \simeq \int_\k k f_\k$ or in terms
  of the fields as
  $N_{\rm dof} V \int_\k \half (E_k^2 + B_k^2)$. Since $E_k^2 \simeq B_k^2$ for
  such plasmons, and $B_k^2 \simeq k^2 A_k^2$ in Coulomb gauge,
  we can also write the energy as
  $\int_\k E_k^2$ or $\int_\k k^2 A_k^2$.  Comparison of these expressions
  leads to the identification (\ref{eq:fdef}).
}
\begin{eqnarray}
f_A(k) & \equiv &
    \frac{k}{N_{\rm dof}V} \langle \A^2(k) \rangle \, , \nonumber \\
f_E(k) & \equiv &
    \frac{1}{N_{\rm dof}kV} \langle \E^2(k) \rangle \, .
\label {eq:fdef}
\end{eqnarray}
Here, $V$ is the total spatial volume and
\begin {equation}
   N_{\rm dof} = 6
\end {equation}
accounts for the two transverse polarization
states and the $3=N_{\rm c}^2-1$ adjoint color states in SU(2)
gauge theory.

On scales where the fields are nonperturbatively large, these
distribution functions are difficult to interpret, depending somewhat on
the gauge fixing procedure.  We do not expect $f_A$ to equal $f_E$;
indeed, if the dominant fields are slowly evolving or unstable magnetic
fields, we expect $f_A > f_E$, perhaps by a large margin.  On scales,
presumably at larger $k$, where the fields are perturbative, we should
see $f_A \simeq f_E$ if the relevant degrees of freedom are behaving as
plasmons with $k>m$.  Therefore, the ratio $f_A/f_E$ serves as a
diagnostic of whether the physics is nonperturbative and whether the
degrees of freedom are primarily independently
propagating plasmons or something else,
such as the magnetic fields associated with hard particle currents.

\begin{figure}
\centerline{\epsfxsize=0.48\textwidth\epsfbox{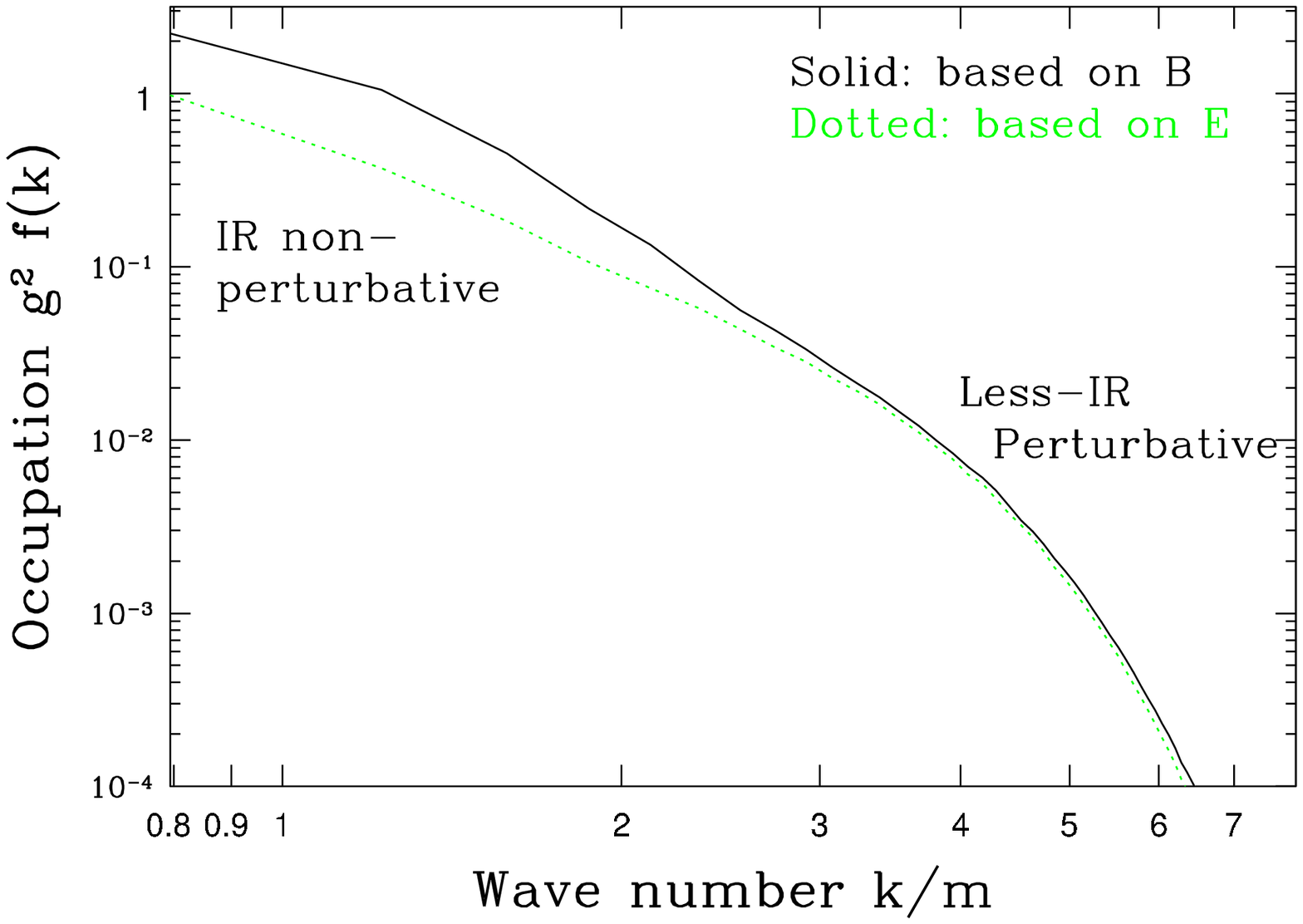} \hfill
\epsfxsize=0.48\textwidth\epsfbox{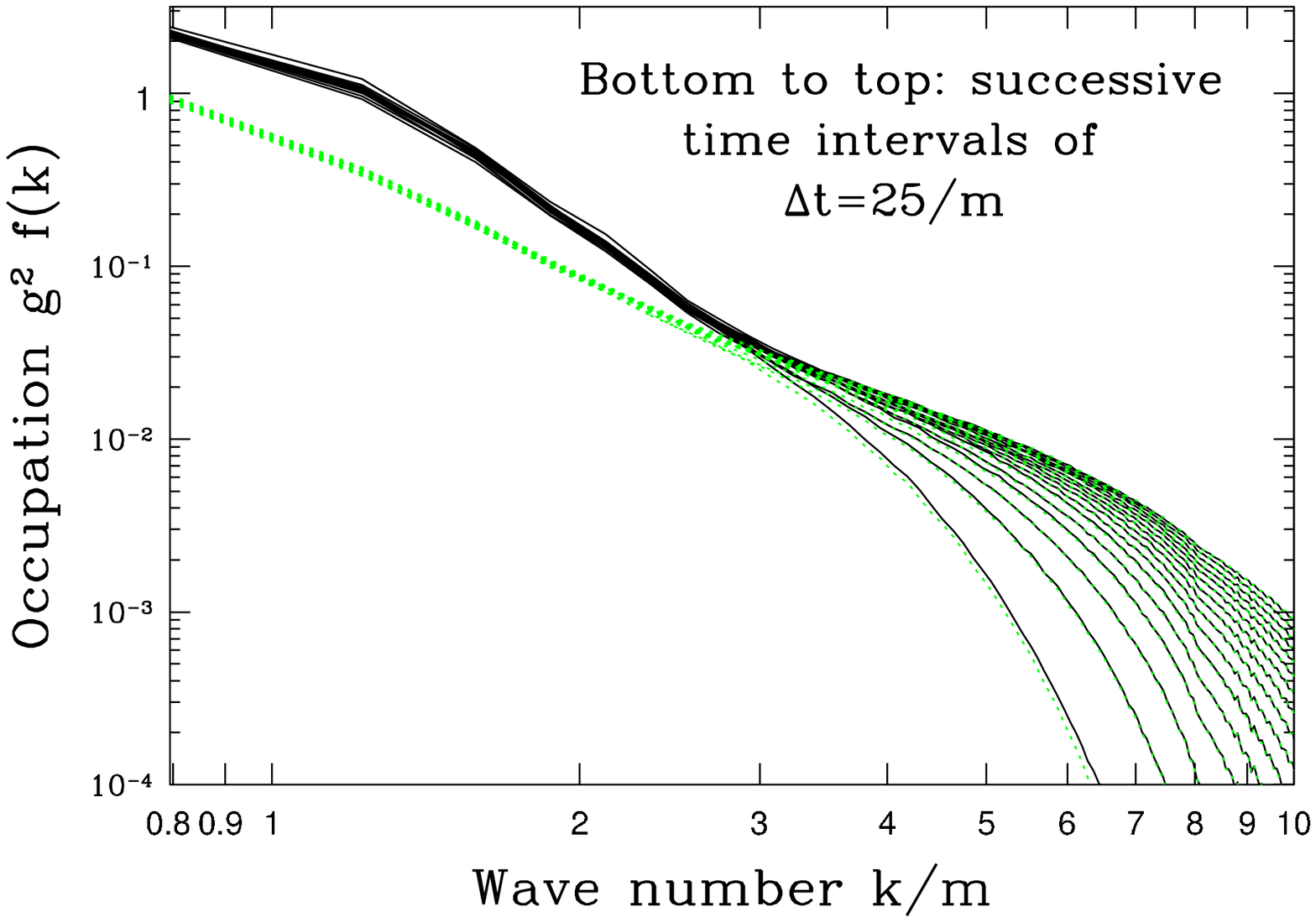}}
\caption{\label{fig:coulomb1} (color online)
Soft gauge field power spectra: {\em left},
initial; {\em right}, as a function of time.
The IR fields are in a
quasi-steady state, and the energy cascades towards more UV modes.
($64^3$ lattice, $am=0.25$, $\lmax=15$)
}
\end{figure}

To test this, we evolved the system for a time of $mt=400$ in a $64^3$
box with $\lmax=15$ and lattice spacing $am=0.25$, corresponding to a
(large) physical volume of $(16/m)^3$, tracking the distribution
functions after the initial transient had died and the magnetic field
energy was undergoing linear growth.  The occupancy after initial
transients is displayed on the left in Fig.\ \ref{fig:coulomb1}, showing
that, as expected, the IR has nonperturbatively large fields, while
fields are perturbative at larger wave number.  On the right in the
figure, we show the time development.  Each curve is time-averaged
over an interval of $\Delta t = 12.5/m$,
and the central times of consecutive curves
are spaced apart by $25/m$.
The IR occupancies
remain nonperturbative but with stable amplitude, while the UV occupancy
increases.  At any $k$, the occupancy rises and eventually saturates;
the saturation point progresses to larger $k$.  This looks like a
momentum-space (kinetic) cascade.

Cascades of energy from the infrared
to the ultraviolet are familiar from turbulence in hydrodynamics and
from many other physical systems.
There are also weakly coupled examples such as weak plasma turbulence
in traditional plasma physics \cite{tsytovich},
and theoretical studies of
post-inflationary thermalization in the early universe
\cite{inflation_cascade}.
Such cascades typically lead to a steady-state, power law distribution
for the power spectrum $f(k)$, usually referred to as a Kolmogorov
spectrum (in honor of the application to hydrodynamic turbulence).
Different microphysics leads to
different powers of $k$.  A thermal spectrum is $f(k)\propto k^{-1}$.
Cascades in scalar field theories during ``preheating''
after inflation, for example, typically display a power spectrum
with various power laws at different stages, such as
$f \propto k^{-3/2}$ and $k^{-5/3}$ \cite{inflation_cascade}.
Obviously we do not expect power behavior at values of $k$ where the
field is nonperturbative; indeed, it is not even clear whether to use
$f_A$ or $f_E$ in this region.  However, we do expect power behavior for
$k$ large enough that $f_A \simeq f_E$, but small enough and at late
enough times that $f(k)$ has become nearly time independent.

\begin{figure}
\centerline{\epsfxsize=0.49\textwidth\epsfbox{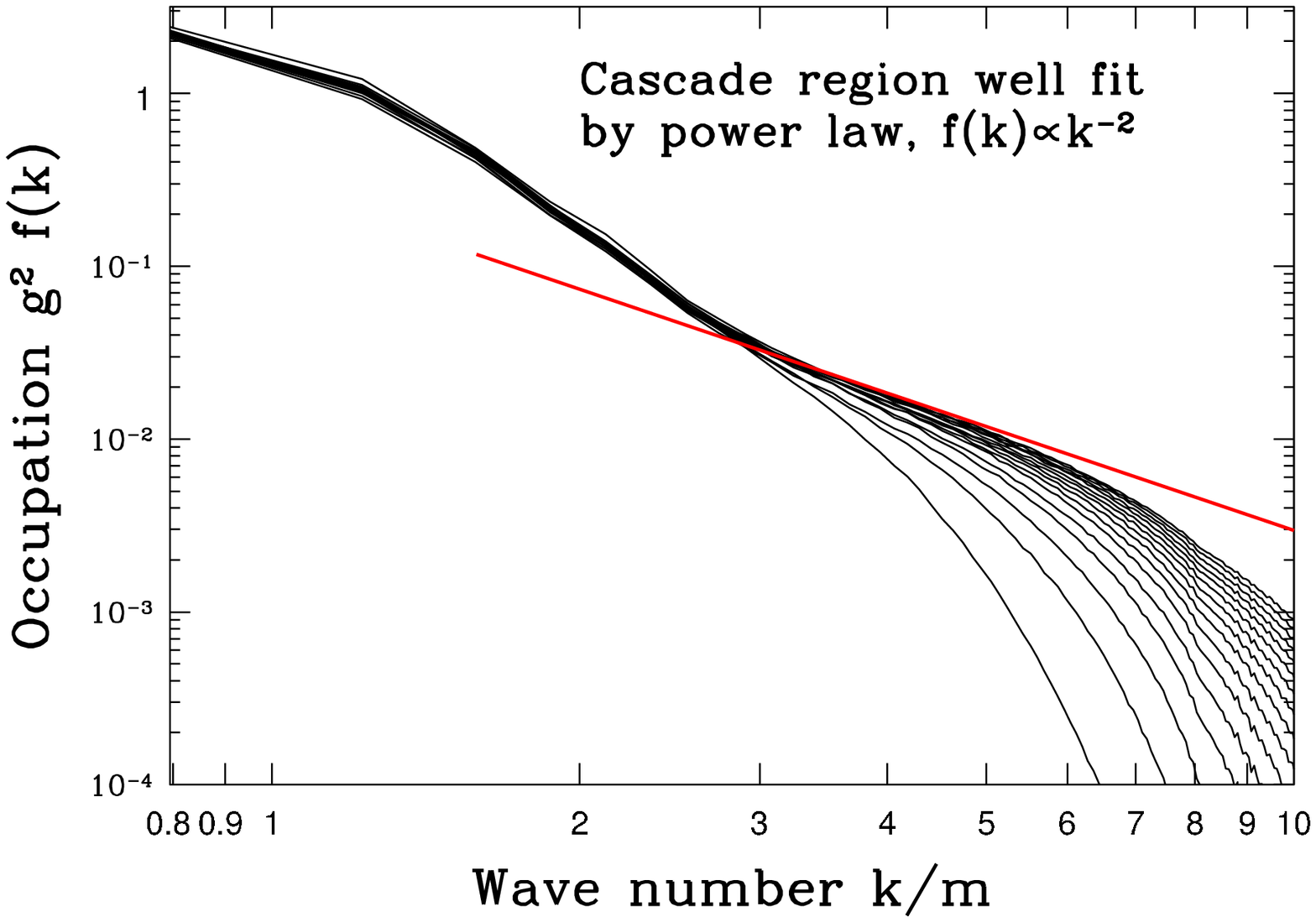} \hfill
\epsfxsize=0.49\textwidth\epsfbox{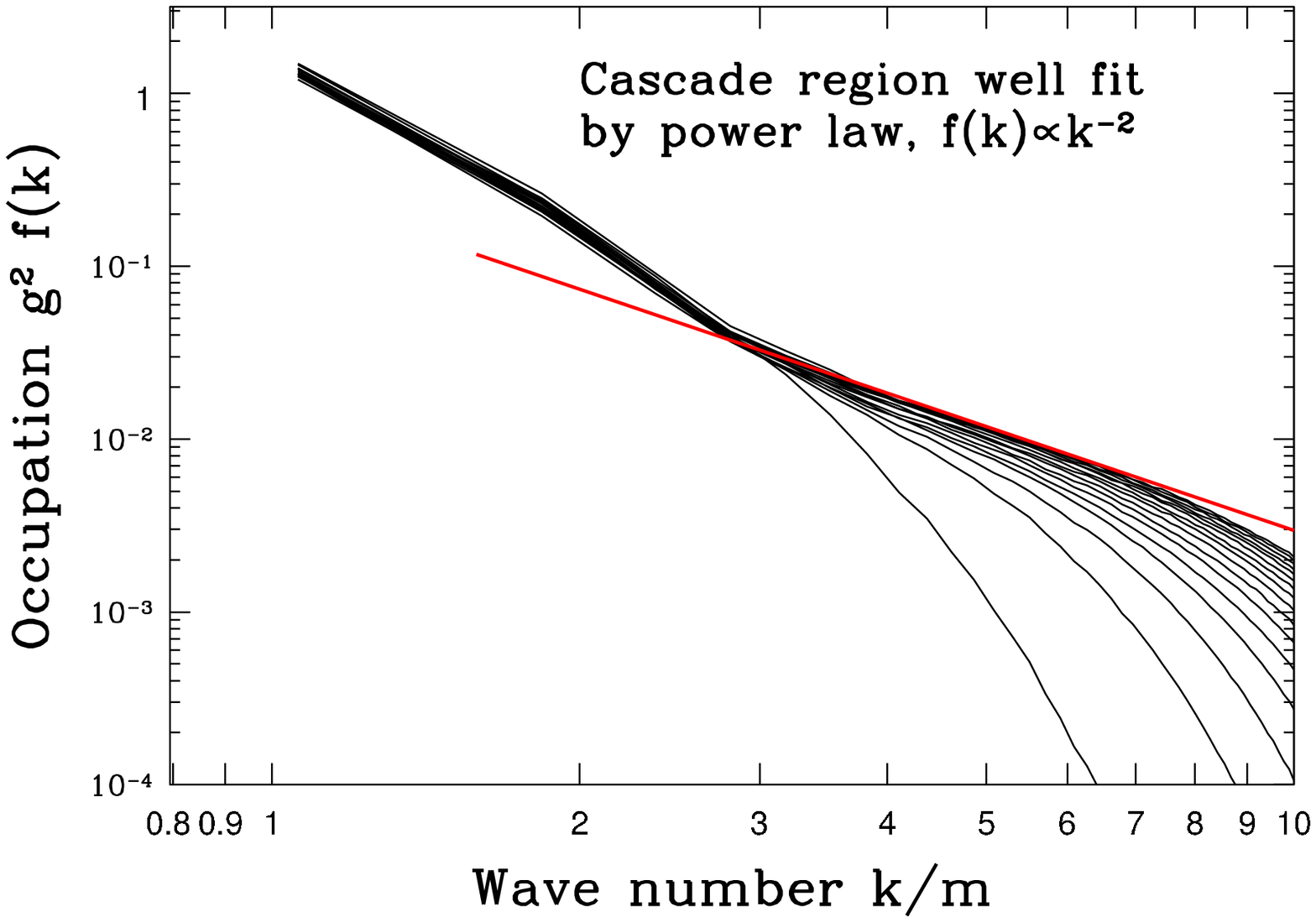}}
\caption{\label{fig:coulomb2} (color online)
{\em Left:}  power spectrum as in Fig.\
\ref{fig:coulomb1}, with a slope $k^{-2}$ power law superposed to guide
the eye.  {\em Right:}  same, but using a finer lattice spacing $am=1/6$ and
twice as much physical time.}
\end{figure}

The first figure in Fig.\ \ref{fig:coulomb2} repeats the righthand
figure from Fig.\
\ref{fig:coulomb1} and shows that the cascade region is well fitted by
a power law with $f\propto k^{-2}$.  Unfortunately, the
most ultraviolet wave numbers involved in the cascade are already at
large enough $k$ that lattice spacing effects may be a concern.%
\footnote{
  For example, there is a small ripple in the first figure of
  Fig.\ \ref{fig:coulomb1} at $k/m=8$.  For $am=0.25$,
  this is the lowest $k^2$
  value where the lattice group velocity can vanish,
  corresponding to a Van Hove singularity.
}
As a
check on the robustness of the result, we performed a second evolution,
also in a $64^3$ box, with the same choice of $\Omega(\v)$, but with
$am=1/6$ rather than $1/4$, and going out to a time of $mt=800$.  The
smaller spacing means that the physical volume was somewhat smaller.
Nevertheless, it was large enough: we have checked agreement within
errors of the $\NCS$ diffusion coefficients, and close agreement in the
energy growth rates.  In any case the physics of the cascade is
presumably more ultraviolet than the physics of the instability, and
should not show severe volume sensitivity.  The power spectrum from this
evolution is shown on the right in Fig.\ \ref{fig:coulomb2}, and is in
very good agreement with the larger volume figure.  The line superposed
is drawn to guide the eye and is not actually a fit; it is precisely the
same line in each figure.  

\begin{figure}
\centerline{\epsfxsize=0.60\textwidth\epsfbox{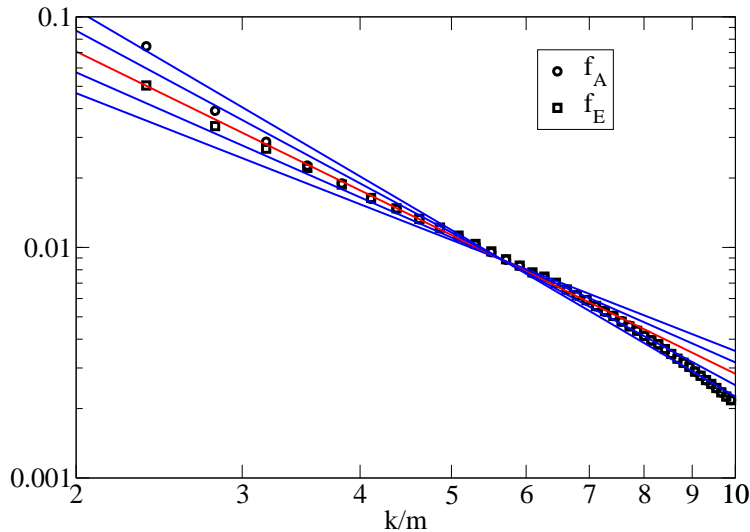}}
\caption{\label{fig:errors} (color online)
The power-law region of the late-time Coulomb gauge distribution $f$ shown
fitted with lines corresponding to powers
$\nu$ = 1.6, 1.8, 2.0, 2.2, and 2.4.
The distribution functions are from the long-time, finer-lattice
simulation shown in the right-hand plot of Fig.\ \ref{fig:coulomb2},
averaged over $700 \le mt \le 800$.  Each data point
represents the center of one of the bins in $k^2$ used to construct
averages in (\ref{eq:fdef}).
($64^3$ lattice, $am=1/6$, $\lmax=15$)
}
\end{figure}

We have found power-law fall-off
$f(k) = k^{-\nu}$ with spectral index $\nu \simeq 2$.
To get a crude idea of the error in our determination of $\nu$, we show
a more detailed view in Fig.\ \ref{fig:errors} of the late-time
distribution on the finer lattice.  We show least-square fits
of various power laws ($\nu$ = 1.6, 1.8, 2.0, 2.2, 2.4)
to the data points%
\footnote{
  We fit the $f_{\rm E}$ data.
  There is only a slight difference between $f_{\rm E}$ and $f_{\rm A}$,
  which is at the IR end of the range chosen.
  The reason for slightly preferring
  $f_{\rm E}$ over $f_{\rm A}$ will become obvious in Sec.\
  \ref{sec:cool} below.
}
in the range $3 < k/m < 8$.
The $\nu = 2$ fit has the longest span of agreement, and we take our
final result to be $\nu = 2 \pm 0.2$.

\subsection{Gauge-invariant cooling}
\label{sec:cool}

We argued that the Coulomb gauge spectrum of Fig.\ \ref{fig:coulomb2}
should be trustworthy away from the IR because the
higher-momentum components of the field are perturbative (as can be
seen from the figures by the drop of occupation number with increasing
momentum).  However, in order to be sure that results are not artifacts
of gauge fixing, it is usually preferable to investigate gauge
theories with gauge-invariant observables.  It is possible to
probe aspects of power spectra in a gauge-invariant way by calculating
the energy of smeared (also known as cooled) gauge fields.
Smearing is a gauge-covariant process which is a function of a parameter
$\tau$ known as the smearing depth.
Define $\A(t,\x;\tau{=}0)$ to be the actual gauge-field configuration
$\A(t,\x)$ at a given physical time $t$.
Then evolve in $\tau$ according to
\begin{equation}
\frac{d A_i}{d\tau} = -\frac{\partial (B^2/2)}{\partial A_i}
	= D_j F_{ji} \, .
\end{equation}
Here $D$ is the covariant derivative and $F$ is the field strength, also
using the smeared field
$\A(t,\x;\tau)$.  This is a gauge-invariant procedure which has a
straightforward lattice implementation.
Such smearings have a long history in lattice gauge theory studies;
for instance they are also used extensively in our
technique for measuring $\NCS$.  Other fields can also be smeared.  For
instance, we define a smeared $W$ field through,
\begin{equation}
\partial_\tau W(\x,\v;\tau) = D^2 W(\x,\v,\tau) \, ,
\end{equation}
where again $D^2$ is the covariant derivative using the smeared gauge
field at the same smearing depth $\tau$.  Note that we do not introduce
smearing into the dynamical evolution of the fields in time
($t$);
we only use smearing for making measurements.
To answer the question ``what is the
smeared $B^2$ at time $t$?'', we make a copy of the fields at
time $t$, apply smearing, and measure $B^2$ on it.

\begin{figure}
\centerline{\epsfxsize=0.46\textwidth\epsfbox{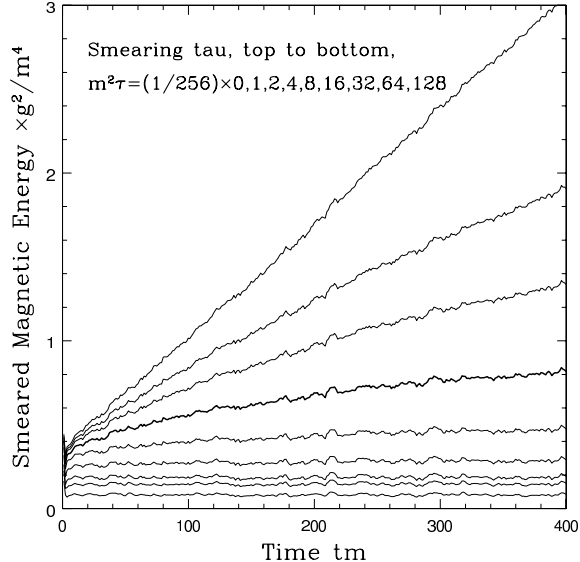}}
\caption{\label{fig:smear1}
Magnetic energy density as a function of time and smearing depth.
($64^3$ lattice, $am=0.25$, $\lmax=15$)
}
\end{figure}

Smeared fields are good at telling us whether the energy going into soft
electromagnetic fields is appearing in very long wavelength,
nonperturbative fields, or in plasmons with larger wave number.  To
study this, we consider the magnetic field energy density $B^2/2$ as a
function of time and of smearing depth, shown in Fig.\
\ref{fig:smear1}.  Very roughly, smearing to a depth $\tau$ eliminates
fields with $k^2 > (2\tau)^{-1}$ and leaves fields with smaller $k$.  But near
$k^2 \sim (2\tau)^{-1}$, the fields are only partially removed, so the real
story is slightly more complicated; smearing is similar to a Laplace
transform of the power spectrum with $k^2$ playing the role of time.
The figure shows that the infrared energy is stable
through the evolution.  (It bounces around, which shows that the fields
are evolving and that it is a small number of degrees of freedom
contributing to the infrared energy.  This bouncing would average out
in larger volumes because of incoherent averaging.)  The total energy
($\tau{=}0$) rises linearly.

Based on the cascade picture of
Fig.\ \ref{fig:coulomb2}, what behavior should we have expected
for the intermediate case of moderate $\tau$?
For any fixed smearing $\tau$, the energy should eventually stop growing
once the modes with $k^2 \lesssim \tau$ have grown to reach their
steady-state distribution in the cascade.  The smaller $\tau$, the
longer it should take to reach the steady state.
The $m^2\tau = 4/256$ curve in Fig.\ \ref{fig:smear1} is
a good example of an initial rise in energy that then tapers off and
is plausibly approaching a steady-state value.  It is unclear from
our data whether the smaller (non-zero) $\tau$ curves will eventually reach
steady-state values.  The problem is that cooling does not select out
a single $k$ but gives a superposition in the form of a Laplace
transform.  Note, for instance, that the $k^{-2}$ behavior of the
Coulomb spectrum of Fig.\ \ref{fig:coulomb1} is limited to the
relatively narrow range of $3 \lesssim k/m \lesssim 5$.  With infinite
computing resources, one could run long enough, on fine enough lattices,
to extend this region over a huge range of $k$, and then one would
expect to see the predicted behavior in Fig.\ \ref{fig:smear1} for
a wide range of $\tau$.

\begin{figure}
\centerline{\epsfxsize=0.46\textwidth\epsfbox{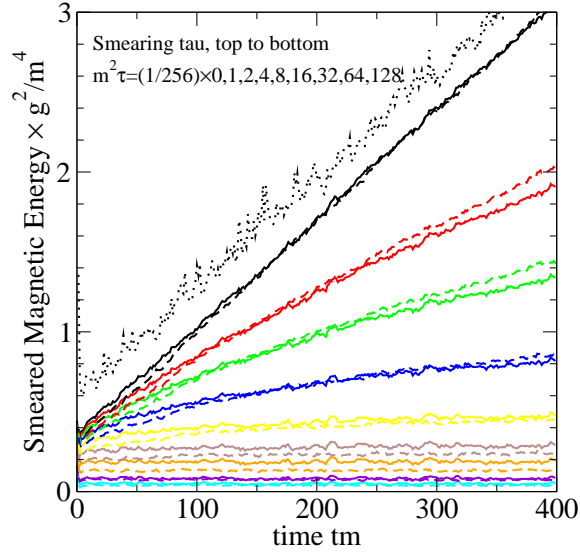}}
\caption{\label{fig:compare}
(color online)
The solid lines are the same as Fig.\ \ref{fig:smear1}.
The dashed lines represent the corresponding results extracted
from the Coulomb-gauge spectra $f_{\rm E}$ of Fig.\ \ref{fig:coulomb1}.
The single dotted line at the top shows a similar extraction of the unsmeared
($\tau{=}0$) curve from $f_{\rm A}$ nstead of $f_{\rm E}$.
The noisy difference with the
corresponding $f_{\rm E}$ curve is an IR effect, and this difference
remains until one cools substantially (not shown).
($64^3$ lattice, $am=0.25$, $\lmax=15$)
}
\end{figure}

But there is a way to check that our Coulomb gauge
results are trustworthy.  We can Laplace transform the Coulomb gauge
spectrum and see if it agrees with the (gauge-invariant)
smeared measurements of Fig.\ \ref{fig:smear1}.  Perturbatively,
the smeared magnetic energy density should be related to the distributions
$f(k)$ by
\begin {equation}
  {\cal E}_B(\tau) =
  \frac{N_{\rm dof}}{2} \int \frac{d^3k}{(2\pi)^3} \,
  k \, f(k) \, e^{-2 k^2 \tau}
  .
\label {eq:compare}
\end {equation}
Note that the growth of $d^3k \times k \sim k^3 \, dk$ with $k$ typically
more than compensates
for the fall of $f(k)$ with $k$ in Fig.\ \ref{fig:coulomb2}, so that
these integrals are dominated by $k^2 \sim \tau^{-1}$ at late times $t$.
We replot Fig.\ \ref{fig:smear1} in Fig.\ \ref{fig:compare}, superposed
with dashed lines that show the results of (\ref{eq:compare}) with
$f=f_{\rm E}$.
The results are very close until one cools deep into
the infrared.
(We also show one sample curve based on $f_{\rm A}$, which gives slightly less
accurate results than $f_{\rm E}$.  The $f_{\rm A}$ have a more
significant IR
contribution than $f_{\rm E}$, as can be seen in Fig.\ \ref{fig:coulomb1}.)
We conclude that Coulomb-gauge
distributions provide a reasonably accurate description of the physics
of the cascade far from the IR, supporting our conclusions based on
Figs.\ \ref{fig:coulomb1} and \ref{fig:coulomb2}.

\begin{figure}
\centerline{\epsfxsize=0.46\textwidth\epsfbox{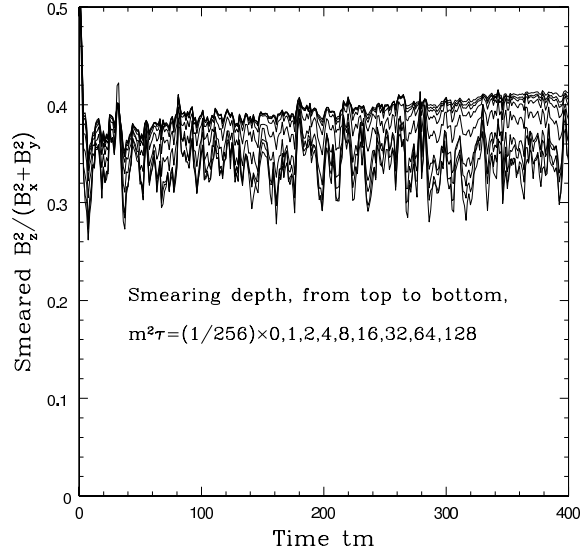}}
\caption{\label{fig:smear2}
The ratio $B_z^2/(B_x^2{+}B_y^2)$ as a function of time and
smearing depth.  ($64^3$ lattice, $am=0.25$, $\lmax=15$)}
\end{figure}

The instability preferentially excites gauge fields with $\k$ vector along
the $z$ axis (the axis about which the particle momentum distribution is
oblate) \cite{ALM,RS}.
Such modes have primarily transverse magnetic fields, so
$B_z^2 \ll B_x^2{+}B_y^2$ for the fields excited by the instability.
Therefore we might expect this behavior of the infrared gauge fields.
If the higher momentum fields represent a nearly thermalized bath, they
will be close to isotropic and the ratio $B_z^2 / (B_x^2{+}B_y^2)$
will be 1/2.  But if they
scatter predominantly off the IR fields, they may also carry a momentum
space anisotropy.  To study this, Fig.\ \ref{fig:smear2}
shows the ratio $B_z^2 / (B_x^2{+}B_y^2)$ as a function of
smearing, for the same lattice parameters as in Fig.\ \ref{fig:smear1}.
Indeed, the soft fields are anisotropic as expected.
The unsmeared fields are (on average) less so, but still have a
definite anisotropy along
the $z$ axis.

\begin{figure}
\centerline{\epsfxsize=0.6\textwidth\epsfbox{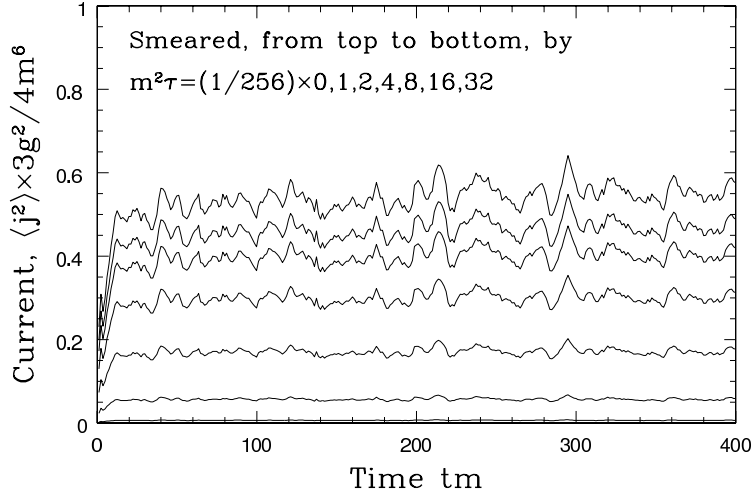}}
\caption{\label{fig:jsmear} Current squared as a function of time and
cooling depth.  The current remains predominantly infrared, and shows
large fluctuations about a nearly flat trend through the evolution.
($64^3$ lattice, $am=0.25$, $\lmax=15$)}
\end{figure}

\begin{figure}
\centerline{\epsfxsize=0.6\textwidth\epsfbox{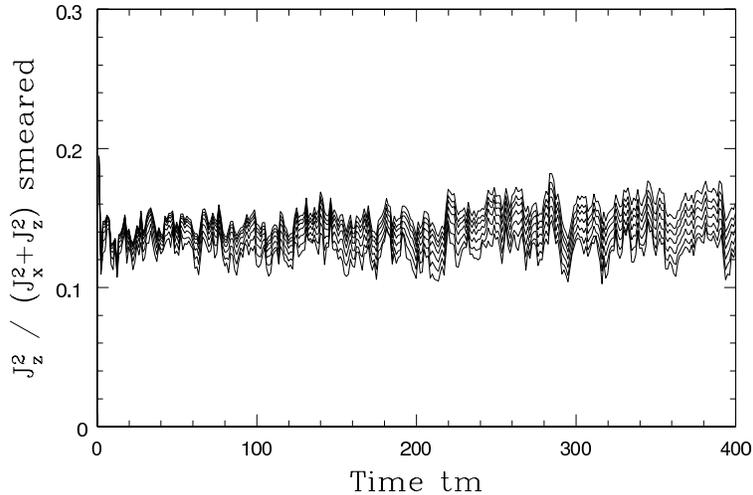}}
\caption{\label{fig:jzsmear}
The ratio of $j_z^2$ to $j_x^2 + j_y^2$ for the various cooling depths
of Fig.\ \ref{fig:jsmear}.  Note the scale of the vertical axis.
($64^3$ lattice, $am=0.25$, $\lmax=15$)}
\end{figure}

The picture that has emerged is that there are soft, anisotropic,
nonperturbatively large gauge fields with higher wave-number plasmons
superposed.  The size of the soft gauge fields fluctuates about a steady
mean, and the plasmons become more numerous, populating higher
and higher wave numbers.  One
way to check whether the interpretation of the high wave number
excitations as plasmons is correct, is to look at the $W$ fields.  We
will concentrate on the $\ell=1$ component of $W$, which is the same as
the particle current up to a factor: $j^2=(1/3) \sum_m W_{1m}^2$
\cite{linear1}.  A plasmon with $k \gg m$ carries almost all its energy
in $E^2$ and $B^2$, roughly equipartitioned, and only a subdominant
amount in currents.  Therefore, if the energy growth really represents a
growing number of plasmons with $k >m$, then $\langle j^2 \rangle$
should grow slowly if at all, and should be IR dominated.
Fig.\ \ref{fig:jsmear} shows our measurements, and the lack of growth
with time is clear.%
\footnote{
   In contrast, a slight growth of current can be seen at very late times in
   Fig.\ 11 of Ref.\ \cite{linear1}.
   This growth appears to be a late-time artifact of the undamped
   treatment of $\lmax$ in that reference, as we discuss in the appendix.
}
[Following Ref.\ \cite{linear1}, we have normalized the curves by
plotting $(3/4m^2) j^2 = (1/4m^2) \sum_m W_{1m}^2$, which is the
contribution of the $\ell{=}1$ components of $W$ to what would be a
conserved energy
$(E^2+B^2)/2 + (1/4m^2) \sum_{\ell m} W_{\ell m}^2$
for isotropic systems.]
$j^2$ falls more slowly with cooling
depth than the magnetic energy of Fig.\ \ref{fig:smear1}, indicating
less power in the ultra-violet.
We also find, in Fig.\ \ref{fig:jzsmear}, that $j_z^2
\sim 0.13 (j_x^2{+}j_y^2)$, nearly independent of time and smearing
depth.  A small ratio of $j_z$ to $j_x$ and $j_y$ is
expected if the currents are primarily associated with
long wavelength, transverse modes in the directions which are
perturbatively unstable.

\section{Discussion and conclusions}
\label{sec:conclusions}

Putting our numerical results together, the physical picture which
emerges is the following.  Plasma instabilities drive IR gauge fields
with $k_z\gg k_\perp$ to grow.  Nonperturbatively strong interactions
between these soft field modes remove energy from these unstable modes,
moving it instead into less-IR gauge field
modes.  The size of the soft nonperturbative fields reaches a
quasi-steady state; if it
grows larger, the nonperturbative physics removing energy gets more
efficient, and if it gets smaller, the instability drives it back up.
The energy absorbed via the instability from the hard particles thereby
powers a cascade of soft gauge field excitation energy towards the
ultraviolet.  The
cascade has power spectrum $f(k) \propto k^{-\nu}$ with $\nu \simeq 2$.
This is not a thermal spectrum,
as also evidenced by the failure of $B_z^2/(B_x^2{+}B_y^2)$ to
approach $\frac{1}{2}$.
We give a theoretical explanation of the value $\nu=2$ in
another work \cite{kminus2}.

These same instability-powered energy cascades should appear in the
early stages of arbitrarily high-energy heavy
ion collisions.  These cascades will only be a temporary feature
along the path to thermalization and will disappear by the
time the plasma is finally fully thermalized.  We leave to future work
the complete integration of the physics of instabilities into
quark-gluon plasma thermalization at arbitrarily high energies.


\begin{acknowledgments}

We would like to thank Larry Yaffe for useful conversations, especially
in the intial development of this project.
This work was supported, in part, by the U.S. Department
of Energy under Grant Nos.~DE-FG02-97ER41027,
by the National Sciences and Engineering
Research Council of Canada, and by le Fonds Nature et Technologies du
Qu\'ebec.

\end{acknowledgments}


\appendix

\section{Improving simulations by damping high \boldmath$l$ modes}
\label{sec:damp}

In this appendix we argue that the large $\lmax$ limit is achieved more
quickly by applying weak damping on large $\ell$ modes than by not doing
so, and we present numerical evidence that this is the case.

\begin{figure}[t]
\centerline{
\epsfysize=7.5cm\epsfbox{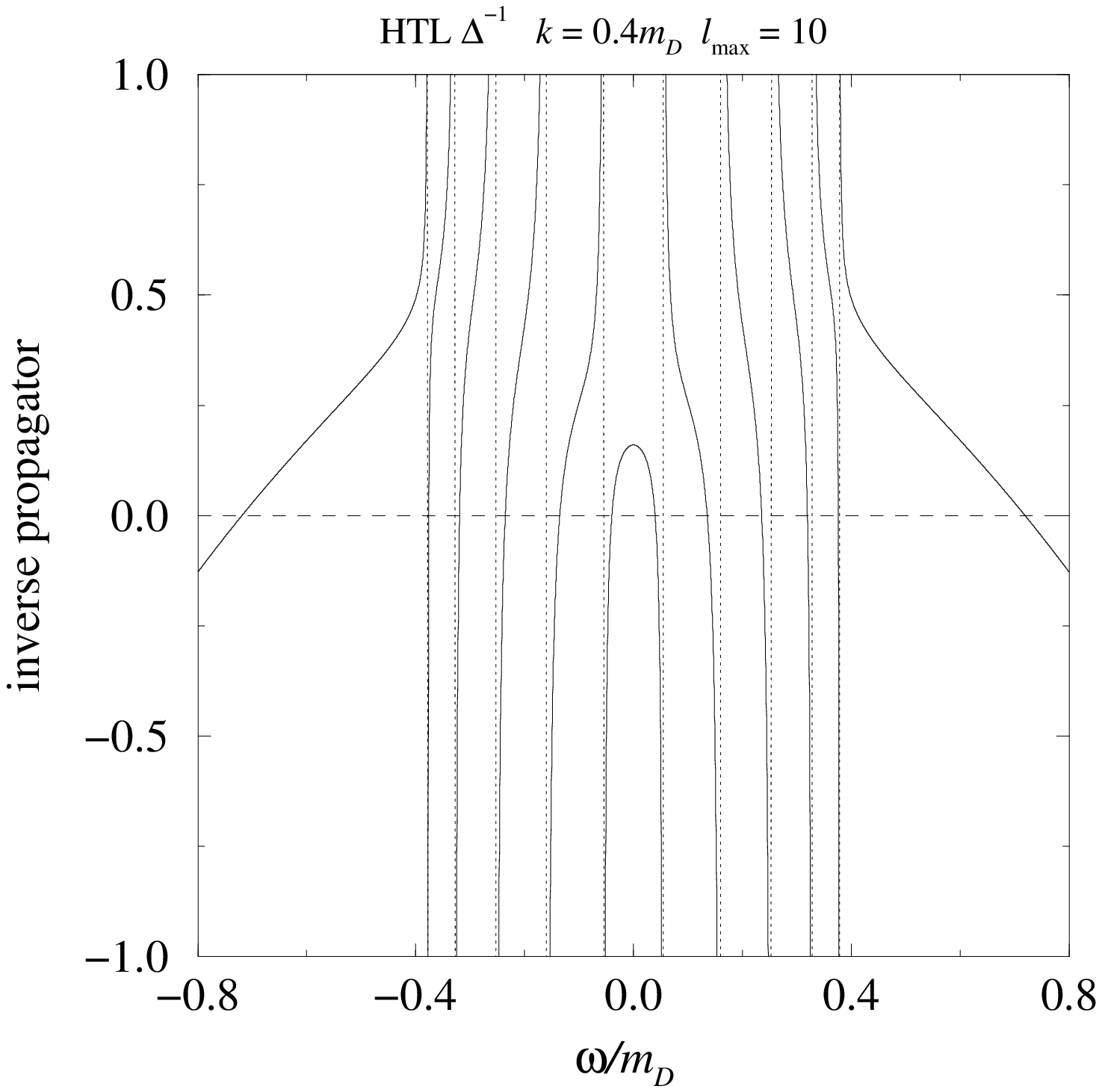}\hspace{4mm}
\epsfysize=7.5cm\epsfbox{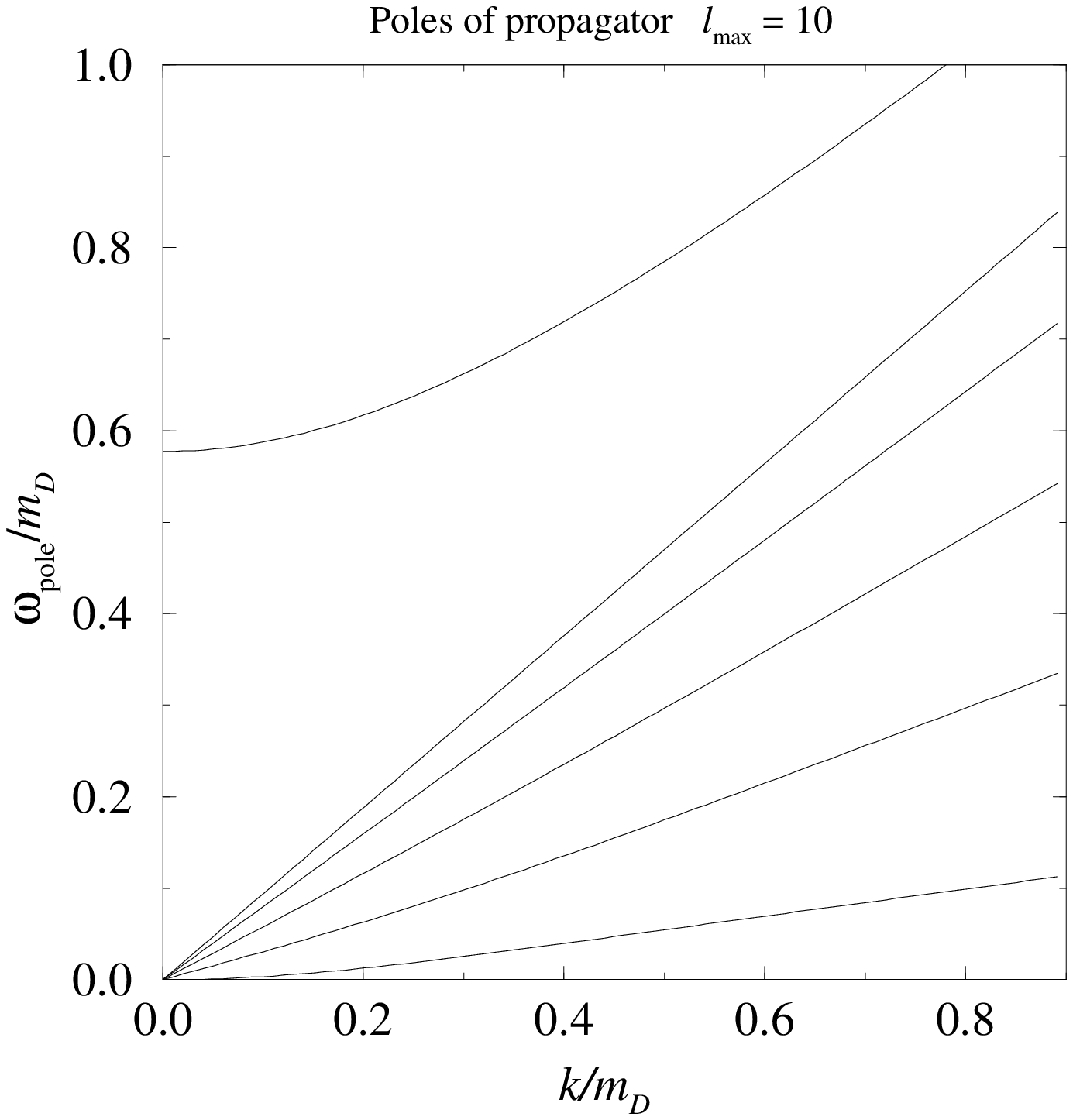}}
\caption[a]{{\em Left:}
The inverse propagator with $\lmax=10$, plotted
against $\omega/m_{\rm D}$ with fixed $k=0.4 \, m_{\rm D}$.  {\em Right:} The
positive frequency poles of the propagator at $\lmax = 10$.  In these
figures, one can clearly see the development of the cut in the
interval $-k \le \omega \le k$, and the two plasmon poles at $\omega^2
\approx m^2_D/3 + 6k^2/5$.}
\label{fig:propagator}
\end{figure}

At the perturbative level, the behavior of the isotropic version of the
$A$ and $W$ field system has been investigated by B{\"o}deker, Moore, and
Rummukainen \cite{BMR}.  The correct analytic structure of the gauge
field propagator in the presence of hard thermal loops (the infinite
$\lmax$, isotropic theory) is that there should be a propagating
``plasmon'' pole at a frequency $|\omega| > k$, and a cut in the
spectral weight for all frequencies $|\omega|<k$.  Physically, the cut
reflects Landau damping.  It means that most of the energy of a long
wavelength magnetic field should be absorbed by the particle degrees of
freedom represented by the $W$ fields, never to return.
However, the finite
$\lmax$ system is a non-dissipative Hamiltonian system.  Therefore it is
not possible for it to contain cuts in the (leading order) propagator.
Instead, the region which should contain the cut contains a series of
poles and zeros in the propagator, illustrated in Fig.\
\ref{fig:propagator}, which is taken from Ref.\ \cite{BMR}.

What this means is that the excitation energy present in the
magnetic field, which is supposed to be Landau damped away, instead
appears only in periodic oscillatory modes.  Most of the energy is
stored in the $W$ fields, as is supposed to happen under Landau
damping.  However, a fraction of it periodically reappears as magnetic
field energy, an effect which is unphysical.  The size of this effect is
determined by the number of $W$ field modes, $(\lmax+1)^2$, which sets
the heat capacity of these modes to absorb and store excitation energy.
The larger $\lmax$ is, the more effectively the $W$ fields can retain
the energy rather than feeding it back to the magnetic fields.

\begin{figure}
\centerline{\epsfxsize=0.46\textwidth\epsfbox{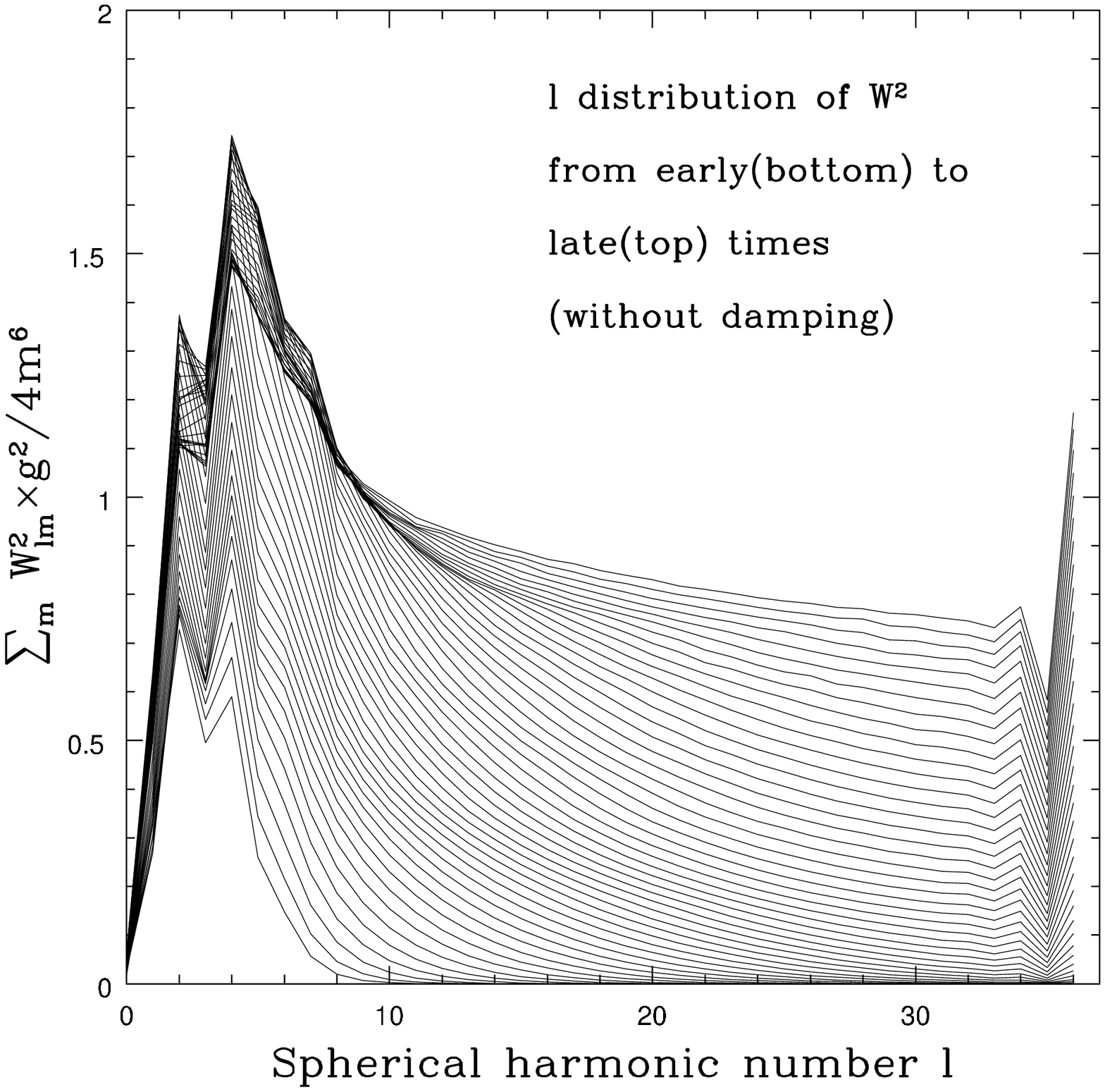} \hfill
\epsfxsize=0.46\textwidth\epsfbox{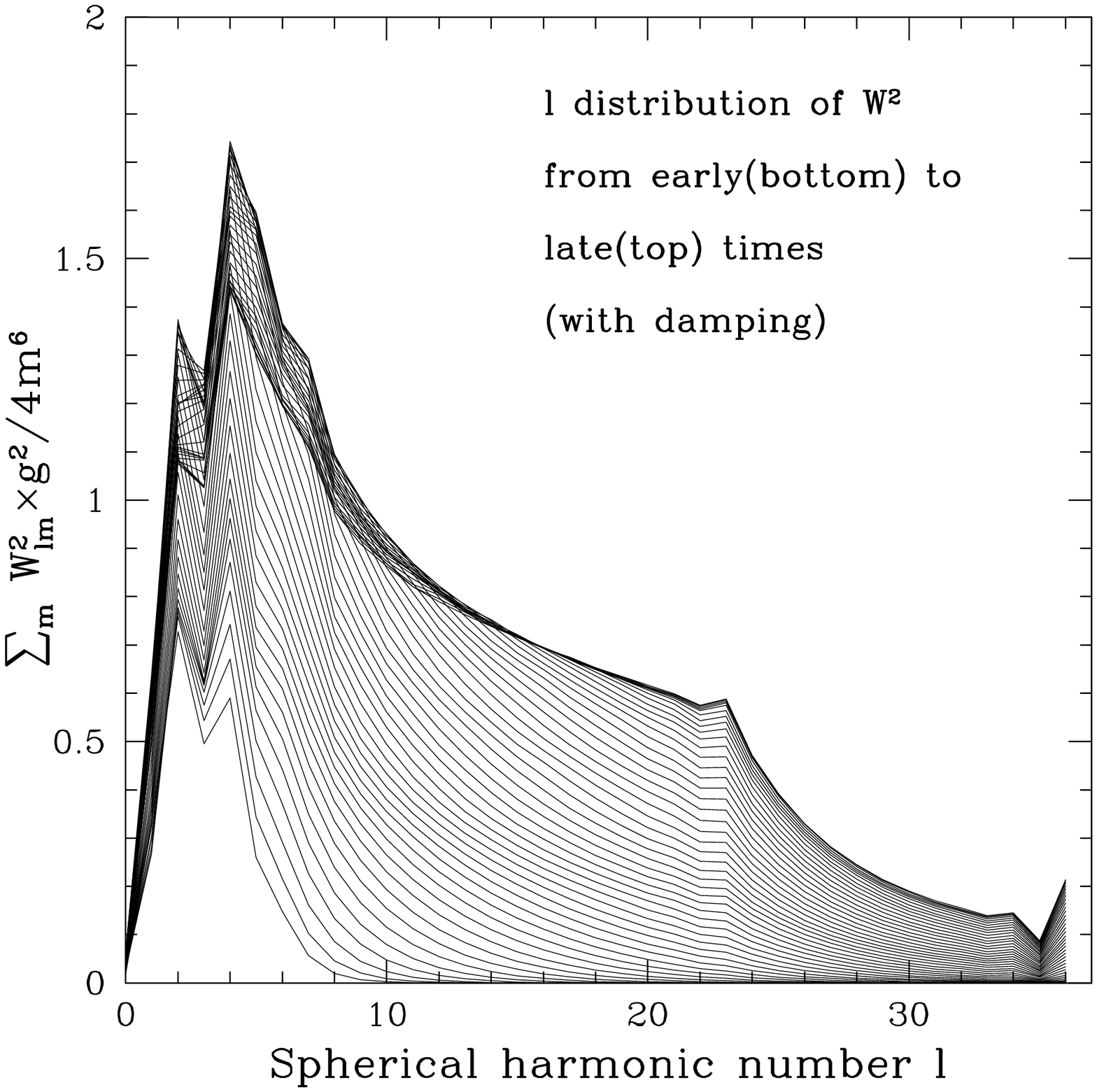}}
\caption{\label{fig:lvalue} Time and $\ell$ dependence of
$\sum_m W_{\ell m}$ the amount of excitation in the $W$ field.  Each
figure plots $\sum_m W_{\ell m}$ against $\ell$ at a series of time
snapshots, for $\lmax=36$.  The left
figure is without damping; the right figure has damping on $\ell \geq
24$ modes.  In each case, the time between successive lines is
$0.57/m$.  The two spikes at small $\ell$ are the angular scales driven
by the instability.  The excitation introduced here cascades to larger
$\ell$, eventually bouncing off the $\lmax$ limit in the left figure but
being absorbed at large $\ell$ by the damping in the right figure.
($50^3$ lattice, $am=0.288$)
}
\end{figure}

The problem in the current context is that, in the anisotropic system,
quite large amounts of $W$ field excitation are generated.  The
derivative term $\v \cdot \D \:W$ in \Eq{eq:W} should cause this
excitation energy to migrate to higher and higher $\ell$ values.  This
is shown in the left-hand plot of
Fig.\ \ref{fig:lvalue}, which shows how $\sum_m W^2_{\ell m}$
varies with time and with $\ell$, if we start with large magnetic fields
but with no initial excitation in the $W$ fields.  However, with a finite
$\lmax$ cutoff, the excitation cannot migrate to arbitrarily large
$\ell$, as it should; some of it instead moves back into lower $\ell$
and re-enters the gauge fields and small $\ell$ $W$ fields.  This causes
a fake increase in the energy of these fields, which becomes more severe
as $\lmax$ is decreased.

Our approach to solve this problem is to assume that any excitation
energy which reaches very large $\ell$ values should continue to
arbitrarily large $\ell$ and be lost to the low $\ell$ system.  We can
make this happen artificially by applying a damping term to the highest
$\ell$ modes, modifying the $W$ equation of motion via
\begin{equation}
\frac{dW_{\ell m}}{dt} = (\mbox{previous}) - \gamma W_{\ell m}
	\Theta(\ell - \ldamp) \, .
\end{equation}
That is, we add a term which causes exponential shrinkage with rate $\gamma$ in
all $\ell$ at or above a cutoff $\ldamp$.  The damping should
only affect the high $\ell$ modes, so we choose $\ldamp = (2/3)
\lmax$.  Note that the damping we apply is gauge invariant: so
long as $\ldamp > 1$, the current remains exactly conserved,
because the current is determined by the $\ell = 1$ modes and the charge
density by $W_{00}$.  We also have to choose a value for $\gamma$.  If
the value is too large, then the $W$ fields with $\ell>\ldamp$
are effectively forced to be zero, which is equivalent to lowering
$\lmax$ to $\ldamp$.  Therefore we should choose $\gamma<m$,
since $m$ is the intrinsic scale of the dynamics.  We should also make
sure we pick $\gamma > m/\lmax$; otherwise the damping is too slow to do
anything, as excitation energy can get from $\ldamp$ to $\lmax$,
reflect, and go back below $\ldamp$ before being damped away.  With this
in mind, we have chosen $\gamma=m/\sqrt{\lmax}$.

\begin{figure}
\centerline{\epsfxsize=0.48\textwidth\epsfbox{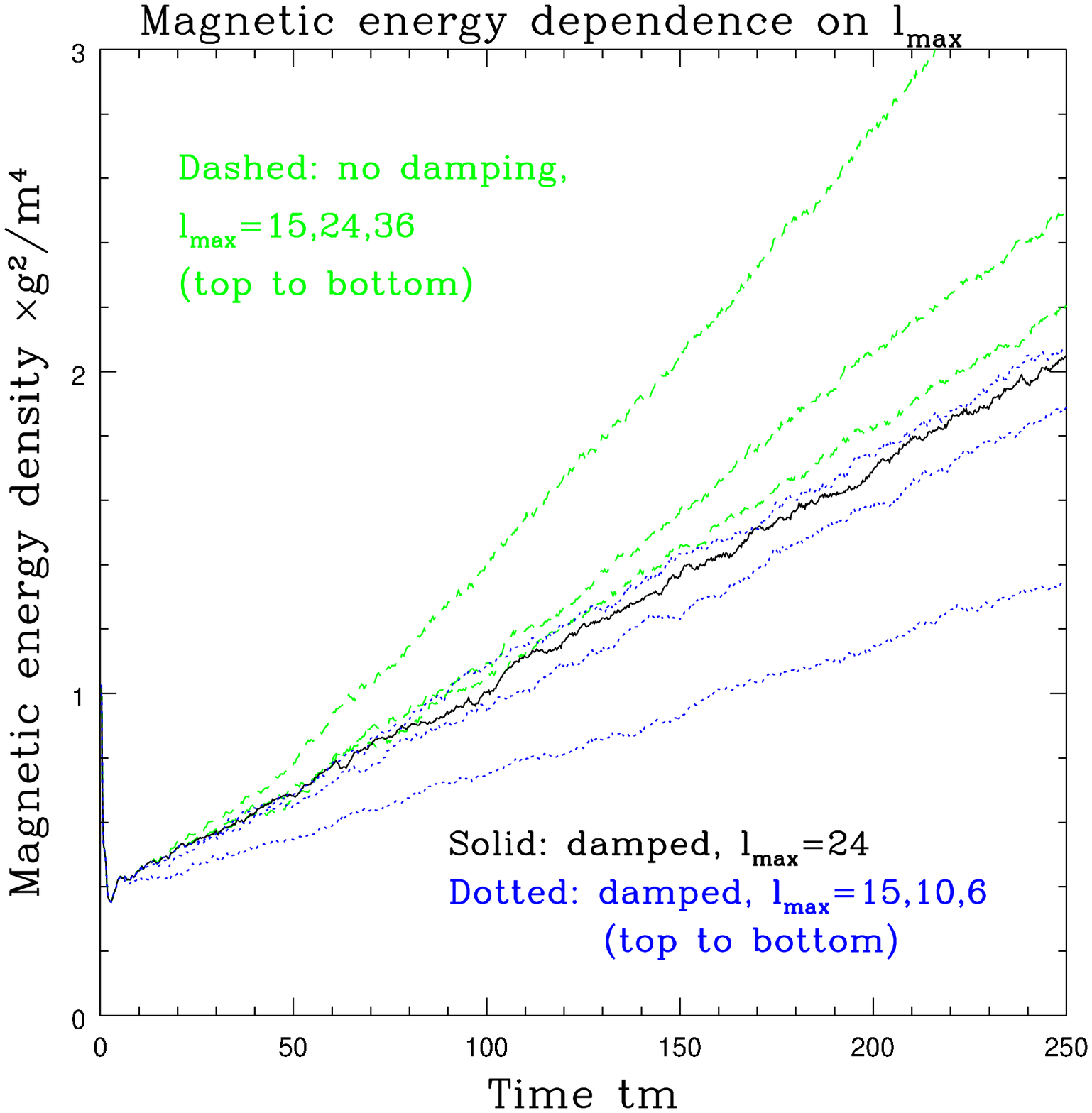} \hfill
\epsfxsize=0.48\textwidth\epsfbox{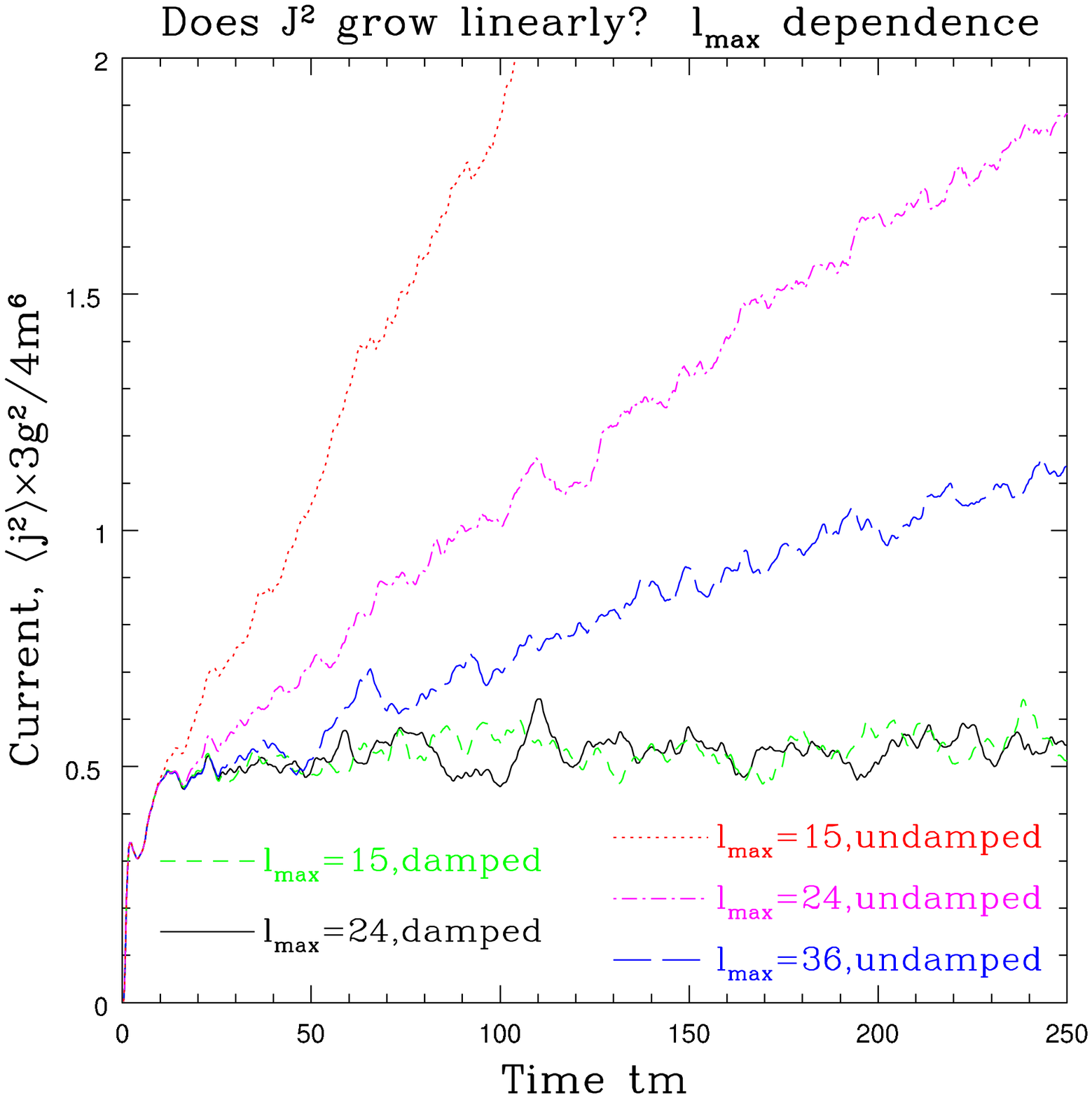}}
\caption{\label{fig:lmax} (color online)
{\em Left:}  magnetic energy against time at different values of $\lmax$, with
and without $W$ damping.  Without damping, the growth rate in $B^2$
approaches a large $\lmax$ value from above, approximately as
$1/(\lmax{+}1)^2$.  With damping, it approaches from below and obtains
the large $\lmax$ value much faster.  {\em Right:} the same for $\langle j^2
\rangle$.  The fact that this quantity does not grow is obtained
immediately with damping, but only approached very slowly
in the large $\lmax$ limit without damping.
($50^3$ lattice, $am=0.288$)
}
\end{figure}

Of course, it remains to test whether this procedure of damping large
$\ell$ modes really gives the same behavior as choosing an extremely
large value for $\lmax$ (which is numerically impractical, since the
number of degrees of freedom rises as $(1{+}\lmax)^2$).  To check, we
have performed evolutions with the same initial random seed and other
parameters  but with different values of $\lmax$, and either with or
without the damping term added.  The results are shown in Fig.\
\ref{fig:lmax} for
$am=.288$, $V=(14.4/m)^3$, and the same $\Omega(\v)$ used in the main body
of the paper \cite{linear1}.
On the left, we see magnetic energy growth vs.\ time.
With and without damping, the growth rates converge
(from opposite sides) towards the same large $\lmax$ limit.
But the damped results converge much faster:
with damping, the magnetic energy growth does not change between
$\lmax=15$ and $\lmax=24$, and appears to coincide with the very large
$\lmax$, undamped behavior.
Now return to the flat result of Fig.\ \ref{fig:jsmear} for the
time dependence of $j^2$.
On the right of Fig.\ \ref{fig:lmax},
we see that this flat behavior
is obtained only very gradually in
the large $\lmax$ limit unless damping is implemented, in which case it
occurs immediately.  For investigating current growth, damping is an
essential numerical tool for practical simulations.

Naturally, if $\lmax$ is too small, we will see errant behavior whether
or not we implement damping.  
For the choice of $\Omega(\v)$ used here
({\it i.e.}\ for the degree of hard particle anisotropy in our
simulations),
Fig.\ \ref{fig:lmax} shows large deviation of the magnetic energy growth
for $\lmax$ below about 10.
We have therefore conservatively used $\lmax=15$
for the bulk of the studies presented in the main text.

\end{document}